# A Trust Framework for Government Use of Artificial Intelligence and Automated Decision Making

# Pia Andrews

*With thanks to the following for your contributions, peer review, comments & support:*
*Tim de Sousa, Bruce Haefele, Matt Beard, Marcus Wigan, Abhinav Palia, Kathy Reid, Saket Narayan, Morgan Dumitru, Alex Morrison, Geoff Mason, Aurélie Jacquet (Ethical AI Consulting)*

As governments increasingly explore and invest in Artificial Intelligence and Automated Decision Making systems, we need to take steps to ensure that these rapidly evolving technologies are used appropriately in the special context of a public service. In many countries, COVID created a bubble of improved trust, a bubble which has arguably already popped, and in an era of unprecedented mistrust of public institutions (but even in times of high trust) it is not enough that a service is faster, or more cost-effective. This paper proposes recommendations for government systems (technology platforms, operations, culture, governance, engagement, etc.) that would help to improve public confidence and trust in public institutions, policies and services, whilst meeting the special obligations and responsibilities of the public sector.

## Table of Contents

# Abstract

This paper identifies the current challenges of the mechanisation, digitisation and automation of public sector systems and processes, and proposes a modern and practical framework to ensure and assure ethical and high veracity Artificial Intelligence (**AI**) and/or Automated Decision Making (**ADM**) systems in public institutions. This framework is designed for the specific context of the public sector, in the jurisdictional and constitutional context of Australia, but is extendable to other jurisdictions and private sectors. The goals of the framework are to: 1) earn public trust and grow public confidence in government systems; 2) to ensure the unique responsibilities and accountabilities (including to the public) of public institutions under Administrative Law are met effectively; and 3) to assure a positive human, societal and ethical impact from the adoption of such systems. The framework could be extended to assure positive environmental or other impacts, but this paper focuses on human/societal outcomes and public trust. This paper is meant to complement principles-based frameworks like [Australia's Artificial Intelligence Ethics Framework](#) and the EU [Assessment List for Trustworthy AI](#).

# Background

Public confidence and trust in "government" is directly impacted by the publics' experience with, and perception of, the public sector. If the service is poor or undignified, if the rules are unclear or inconsistently applied, if policies seem misaligned with public values – this all undermines public trust and confidence in the sector, and in all that the sector administers. From laws and policies to national security and elections, public trust and confidence is critical for a functional, fair and stable society, as has been demonstrated globally throughout the pandemic. For all these reasons, public institutions have always operated within a special context, where all actions and decisions are governed differently to other sectors to ensure accountability and appropriate use of delegated powers.

Trust can be broadly defined as a) a willingness to believe that a person or entity is operating in good faith, b) with integrity, and c) in a way that fulfils the individual's expectations of that person or entity. Government systems could therefore be considered trustworthy through demonstrating **good faith** (through a systemic and measurable commitment to human-centred and humane outcomes), by assuring **high integrity** systems (that are lawful, accurate, high veracity, assured, consistently applied and appealable), and that meets **public expectations** (by reflecting public values and needs, doing no harm, being transparent and operating within relevant legal, social, moral and jurisdictional limitations of power).

Unfortunately, the way in which governments have digitised manual processes across the breadth of public service functions has often resulted in opaque, inscrutable systems that are not explainable or traceable back to their legal authority. For instance, many systems produce outputs without recording the correlating legislative rules, data and other factors the output was based upon, requiring manual review in the case of a challenge. This has made it difficult to understand or appeal decisions made by systems which are not auditable in real time and are incapable of detecting and minimising unintentional harm.

Most systems are not measured in terms of human benefits or impacts, which makes understanding or driving human outcomes impossible. The digitisation of public systems, processes and now decision-making, without the preservation of accountable transparency, has unintentionally (but, perhaps, foreseeably) created several issues for public trust.

The key challenges that have emerged today include the following:

- **Disenfranchising users of government services:** Citizens/residents are increasingly faced with a 'computer says no' situation from public services, with unclear or difficult-to-access avenues for seeking an explanation or appealing.
- **Disconnected service:** Front line staff have little means of determining, let alone explaining, the rationale behind decisions generated from opaque software systems, which in turn are often protected from scrutiny by commercial-in-confidence or copyright vendor arrangements.
- **Garbage in, garbage out:** An increasing number of AI/ADM systems are machine-learning based, using historical data to train responses - this creates two problems:
    1. can perpetuate historical bias/inequities and legacy or system-based mistakes without traceability to law, and displaces the possibility of purposefully equitable service delivery (i.e., 'fairness-by-design'); and
    2. ML based systems differ from rules based ones in that they are **stochastic**, meaning they create additional uncertainty because they are probabilistic and create inconsistent results over time. The same inputs do not yield the same outputs, creating a inherent and unfair contradiction to Rule of Law. Rules-based systems by contrast are **deterministic**, where the same set of inputs consistently give the same output. Systems used in government to create a decision/action that affects a person should be rules-based or at least be tested against rules to ensure the outputs are legally valid.
- **Low traceability to authoritative rules:** Where rules are coded into public sector software systems, including AI/ADM systems, they are currently a mashup of legislation and operational policies, where departmental policies may conflict with, override or undermine legislative rules - this can undermine the veracity and legitimacy of system outputs.
- **Prioritising money over people:** Individual decisions in an ADM system remove humans from the decision-making loop, with a subtle but powerful shift in incentives from human outcomes to financial or efficiency measures, creating an unintended human cost as the system prioritises money/savings over people.
- **Unknown human impact:** Public servants generally try to do no harm, but there is no consistent and repeatable way to measure human impact from government systems, services or programs. This creates a genuine barrier to identifying if and when new policies or changes create (likely unintended) harm, either individually or across the system as a whole. Public services must be committed to the practices of both identifying foreseeable harms, and to *explicitly* and continuously check systems for unintended or unforeseen harms if they occur.
- **Disenfranchised polity:** The public feel (and generally are) unable to easily participate in the development of policies, laws and services that affect them, and are rarely given access to the rationale, data or government models used.
- **Homogenous policy development:** Policy development rarely adopts a shared and culturally inclusive approach from the start of the development cycle, resulting in policies that are not meaningful to or reflective of diverse community values and needs.
- **Disconnected from Country:** Public sectors rarely engage with Indigenous knowledge systems to create policies, programs or systems that are connected or reflective of [Country](Country), and are not building Indigenous values, lore/law or impact measures into government systems.
- **Secretive culture:** The public sector continues to operate on a highly secretive and 'need to know' basis, creating trust issues when citizens or media are unable to reasonably scrutinise systems, processes or operations. This also creates a barrier to access. For instance, in many government systems, the rules for eligibility, entitlement, etc. are not easily publicly available.

> This creates a barrier to people getting the help they need, especially at their most vulnerable. It also makes it more difficult to identify non-compliant systems or processes.

Finally, a key reason for developing this framework is to supplement the necessary, but insufficient, approach commonly taken across the world adopting a principles-based and/or human reviews-based ethics frameworks for AI (or indeed in any system). There is a strong presumption that if qualified humans review a system before, during and/or after implementation, then an ethical outcome will be achieved. This approach has some fundamental challenges:

1. There is usually inadequate training, education about or sufficient resources to properly implement ethics frameworks and principles, often within environments unsupportive of fully using them (often due to the pressure of expediency or top down decision making).
2. The premise that human involvement leads to ethical outcomes seems in stark contradiction to much of human history. "Ethics" are, of course, highly subjective and contextual. As such, what one person considers adequate might be anathema to another, especially in a principles-based framework where interpretation is personal. This is further complicated where there are naturally conflicting interests and/or power imbalances.
3. Ethical panels and review processes usually do not include representation of those actually affected by the system being reviewed. As such, reviewers frequently do not have sufficient experience or perspective to identify or address the full impacts on affected people and communities, and panels rarely invite a diverse range of knowledge systems.
4. If there are no minimum requirements that are prescriptive, measurable or standards-based, the resulting "ethical systems" are likely to be inconsistent with each other, and not measurable against any meaningful human impact criteria. How can you "avoid harm" if you don't measure for harm? How can you ensure accuracy if you don't have authoritative measures to assure against? How can you avoid bias if you aren't testing for consistency?
5. Finally, a principles-based human review approach ignores the fact that people are not well equipped to deal with complexity. Modelling currently done is usually limited to the domain or portfolio of the target policy, which ignores the likelihood, predictability of and measurement of secondary impacts from across the full jurisdictional context.

None of this is to say that ethics frameworks should not include human review, but rather that we need **more** than just human review to consistently ensure ethical outcomes over time. The framework laid out in this paper is intended to **complement** human review frameworks with some practical means and recommendations for government systems, specifically to mitigate the issues laid out above. Any one part is not sufficient to assure high trust and high veracity systems.

## The special context of government requires a different approach

These issues are anathema to good government. Government has a special context that differentiates government from private sector organisations, and requires a different approach to the design and delivery of systems. Building the special context into the design approach would largely address the problems laid out above, and would in turn ensure and assure public outcomes and are fair, accurate and lawful, which would build and maintain public trust, and, by extension, maintain legitimacy.

- **Administrative Law** - a special requirement for public sectors is accountability to the Parliament, people, Auditors General, as well as being auditable and compliant (testable) with legal authorities and the principles of Administrative Law. Access to review of gov decisions is a key component of access to justice - you cannot review what you do not record and explain.
- **Access to justice** - governments have significant power in society, with the ability to penalise, incarcerate, fine, tax, etc. Therefore, services generally need to be explainable and easily appealable and testable against accessible legislation to ensure access to justice.
- **Privacy preserving** - governments have very specific requirements as set out in privacy and data handling legislation (such as the [*Privacy Act 1988* (Cth)](#)) to enable transparency of use, appropriate use and disclosure, and protection of personal information and personal privacy.
- **Human outcomes** - most things done in government are ostensibly in the pursuit of a policy objective and for public good. Measuring and monitoring the ***policy impacts*** over time is critical to assuring that the policy intent is met, but monitoring and measuring the ***human impact*** is also critical to ensuring a net positive impact for society with limited unintended harms. Both types of measurements can feed into policy iteration and responsiveness to change. This is different in the private sector, which is primarily driven by a financial imperative.
- **Human and moral rights** - Governments must comply with relevant human rights, international and domestic. Citizens/residents also have a moral right (even if not enshrined in law) to good, trustworthy, legitimate government. Governments arguably fail when people come to reasonably believe their rights are at risk, or that the public sector don't care, is not acting in the best public interests or cannot be relied on to secure the basic wellbeing of the community. For example, [a 2016 study of 20 countries across sub-Saharan Africa](#), found that perceived corruption reduces wellbeing, but that high trust in institutions can mitigate that impact. Low trust means political and ethical failures are going to be detrimental to our wellbeing. And to the extent that AI/ADM will be used to mediate relationships between gov and citizens, it is certainly a part of the special context.
- **Constitutional and/or legislative purpose** - there is often public/political debate about big or small government, or what government should or shouldn't do. But there is always a constitutional and/or legislative mandate for public institutions, which they need to deliver upon regardless of the particular political debates of the time. This context is important for designing AI/ADM systems that are both compliant and aligned to the purpose and mandate of the specific public institution and jurisdiction, in contrast with commercial systems. The mandate of public institutions ideally drive their short and long term objectives and outcomes, and thus should form the base incentives and purpose of AI/ADM systems.
- **Democratic context** - representing the will of the people and operating with their consent is part of the special context of government and by extension, the public sector. Public sector organisations are meant to serve the Government of the day, the Parliament and the People, making them subject to scrutiny by both the Parliament (and Opposition), and the communities they serve. Also, reduced public confidence in the public sector leads to people simply not trusting, engaging with or respecting as legitimate all the public sector administers, including democratic processes and outcomes. Trustworthy systems are therefore important for helping ensure public trust in democratic outcomes, as administered by the public sector.
- **The state monopoly on violence** - finally, and possibly most importantly, governments have high impact levers that no other sectors have. Government can investigate, penalise, enforce, seize assets, institutionalise, or incarcerate. Strict controls are very important to ensure these powers are not abused and to maintain public confidence. This is also why separation between the executive and judicial branches of government is so important: so everyone, including governments, must be equally able to be held to account under the Rule of Law. While many people may share large amounts of personal data with private sector technology platforms such as Google or Facebook, they have a choice as to which services to use, and what to share, and those companies do not have the power to imprison them, or take their children away. People do not have the choice to not interact with government, and rarely have the choice of which governments to interact with. This context is extremely important because if government systems are not fair, equitable, lawful, appealable, etc, then the real impact on people's lives can be, and has been, devastating.

Examples of the type of human devastation that can be caused by government ADM systems include the Australian Federal Government's *Online Compliance Intervention* scheme (also known as 'Robodebt'), which was the subject of a significant class action settlement worth ~$1.2 billion, in which the Government agreed to refund or cancel 470,000 incorrectly levied debts and pay additional compensation. Notably, there were numerous reports of harms caused by the scheme, in terms of stress and financial hardship, including allegations of suicides.[1]

Another example can be found in the decision in [Connecticut Fair Housing Center, et al. v. CoreLogic Rental Property Solutions, LLC](). In this 2019 matter, a single mother in rental accommodation in the US state of Connecticut was refused permission for her disabled son to move in with her. Her son had been injured in an accident which left him unable to walk, talk or care for himself. The refusal was made on the basis of a decision made by CoreLogic's automated tenant screening software tool, 'CrimSAFE'. The tool did not provide reasons for the decision. It was later revealed that the reason was a shoplifting charge against the son which had been dropped, and which had occurred some years prior to the accident. As a result of the lack of transparency and explainability of the automated decision, the plaintiff's son was required to be housed in a nursing home for a year longer than was necessary, causing stress and financial harm to both the plaintiff and her son.[2]

Governments affect people's lives, ideally for the better, but sometimes for the worse. As governments increasingly digitise systems, government systems, including ADM systems, must be designed to live up to the basic expectation of delivering measurable public good, fairly. And where they do not, mechanisms are needed to identify and take corrective action.

# Trustworthy framework for AI/ADM systems

## Designing for trust

So how can you build trustworthy systems in government? It is useful to clearly articulate the requirements for building and maintaining trustworthy systems in the special context of the public sector. These six questions should be put at the heart of the design of all policies and services, to help make them more trustworthy:

1. How would you audit and monitor the decisions/actions made, their accuracy and their legal authority, in real time?
2. How would an end user (citizen, resident, etc) know, understand, challenge and appeal a decision/action?
3. How would you know whether this action/process is having a fair, positive or negative impact?
4. How would you ensure and maintain independent oversight and effective governance?
5. How would you detect, respond to and implement continuous change, external or internal?

---

[1] https://www.abc.net.au/triplej/programs/hack/2030-people-have-died-after-receiving-centrelink-robodebt-notice/10821272 ;
https://www.sbs.com.au/news/article/mothers-whose-sons-took-their-lives-after-robodebts-detail-anguish-in-heartbreaking-letters/dcxjk7ex2
[2] https://www.cohenmilstein.com/case-study/connecticut-fair-housing-center-et-al-v-corelogic-rental-property-solutions

6. How can you operate in a way the public would consider 'trustworthy'?

Every government system demands a solid answer for **all** of these questions, but the ultimate test is to *ask people* what would make a relationship with a particular agency trustworthy (which will vary according to the mandate of the agency) and implement measures accordingly, rather than assume or demand trust. Trust is situational and contingent to the context of legislation and the scope of the institution that seeks to gain and maintain that trust. Mapping the end to end 'user journey' or process for these questions inevitably forces us to dig into several likely features of a high trust system.

Below are some likely functions or components of a trustworthy system, structured to answer the questions above, noting some of the functions would be required for more than one question.

## 1. How would you audit and monitor the decisions/actions made, their accuracy and their legal authority, in real time?

- **Explainability and decision capture:** Have you recorded the events leading up to the decision, the data and legal basis upon which the decision was based, and the decision itself? The [Artificial Intelligence Ethics Framework](#) for Australia suggests that end users be advised when they are interacting with an AI system for transparency, but transparency in the public sector context also requires transparency in how a decision was made & with what authorities. Explainability of the decision/action itself is *necessary* for public sector ADM systems.
- **Traceability of authority:** Can you tie the process and its outcome back to legislative, regulatory, delegated, or policy authorities? This requires prescriptive legislative and regulatory rules to be digitally available, testable and traceable by machines, and clarity on what rules require human judgement with digital access to case law to draw on precedent.
- **Monitoring for accuracy:** Is there a means to check the consistency, accuracy and predictability of your system outputs? Is the test suite or verification mechanism publicly available for anyone to test against? Are you monitoring for accuracy, and for patterns to identify unusual trends that need to be understood or investigated?
- **Operational responsiveness:** There is no point having a monitoring system if there is no operational model around it with skilled staff able to take action or escalate an issue. A trustworthy system needs to be a combination of people and technology capabilities.

## 2. How would an end user know, challenge, understand and appeal a decision/action?

- **Decision discoverability:** How do you communicate the decision to the affected person? How can they access their record of decisions and trust the record was not changed?
- **Transparency and traceability**: How can the affected person understand the decision: the rules, the evidence that they were applied to, and how that led to the decision?
- **Ease of access to appeal:** How can the end user (citizen, resident, etc.) appeal the decision? What is that 'user journey' and how can you ensure a dignified experience throughout?
- **Consistency of application:** Is the system applying rules and decisions consistently, with consistent outcomes that don't unintentionally discriminate against or bias a person or group?
- **A right to explanation:** The onus must be on the government entity to explain the decision/action, but this explanation is only available through traditional appeals or FOI processes, which are a barrier to explanation. It might be helpful to introduce an explicit *right of access by person to reasons for decisions affecting that person* (as found in [Section](#) 23 of the OIA in New Zealand), creating a legal obligation on departments to explain decisions to the people about which decisions are made.

## 3. How would you know whether this action/process is having a fair, positive or negative impact?

- **Human outcomes measurement framework:** Is your system measuring the human impacts from your AI/ADM system, both directly and broadly? Can you measure the real human impact of change to your service or policy? How can you detect and mitigate unintended harm?
- **Societal based measurement framework:** Is your system measuring or motivating good outcomes for society more broadly?
- **Immediate and long-term timeframes:** Can you investigate the decision in real time, at an individual, system and societal level? Can you model change across whole-of-government?
- **Impact data collection:** Are you collecting relevant data (from your system or from other sources) to understand the impact of your AI/ADM? Is this data protected against misuse? Are you ensuring a privacy-by-design approach, to meet the letter and spirit of the Privacy Act?

## 4. How would you maintain independent oversight and effective governance?

- **A library of models and algorithms for AI/ADM systems:** Can the public find, understand and test the building blocks of your government AI/ADM system? If not, can at least the relevant oversight/governance bodies do so (particularly in the case of auditing agencies and the Parliament, to oversee less transparent systems)? What skills and tools are needed to audit and monitor such systems?
- **End-to-end data/software/algorithm assurance:** Can you test, audit and monitor the full software and data supply chain to ensure no interference in the resulting outputs/decisions, as well as objective validation of data provenance and quality? It's also important to test and monitor how different systems/rules/data interact, to avoid harmful unintended consequences.
- **Participatory governance:** Is there a broad and diverse representation of the society served included in the oversight and scrutiny of the system?
- **Public reporting:** What visibility does the public have on the operations, policies, services and administration? What do they need, to trust that specific public sector organisation? In the same way that data breaches must be reporting, where a system has been identified as having an adverse effect, these should be publicly reported.

## 5. How would you detect, respond to and implement continuous change, external or internal?

- **Detecting and responsive to change:** Are you monitoring for human impacts over time, benefits and harm, and for extrinsic or unexpected changes that would trigger a policy or implementation change? Do you have an operational model that supports continuous evidence-based change and improvements to policies or services?
- **Highly skilled workforce:** Public institutions need a highly skilled workforce that can effectively deliver core functions with integrity, including in this context, AI/ADM systems, monitoring/measurement and auditing. Researchers and vendors can be engaged to support

   capability uplift, and provide platforms/tools as is appropriate, but for public institutions to have public trust, they have to be able to deliver well the core functions they are responsible for. This includes having the skills to both deliver systems internally, and to appropriately specify vendor deliverables and hold vendors to account for delivering them.
●  **Feedback mechanisms:** Is there easy to access means for anyone to provide feedback to your system, including staff and citizens? Is feedback collected throughout the process of delivery a service/decision? Are new insights being continuously created and actioned? Is it monitored, analysed, prioritised and actioned to feed continuous improvement?

## 6. How can you operate in a way the public would consider 'trustworthy'?

The [Australian Privacy Commissioner's report](#) on community attitudes about privacy and trust identifies that the public has some common and increasing concerns: worries about data and privacy protection (including personal location data), to know when their information is used in ADM systems and how that will affect them, access to justice, etc. Generally speaking, when users do not trust a system, they avoid contact with it or route around it. Accordingly, a lack of trust can hobble or wholly comprise the effectiveness of a system, and the relationship between the person/community and the organisation that runs the system. Below are some ideas to consider in designing, delivering and operating a system in a way that the public could consider trustworthy:

- **Participatory administration:** public institutions should continuously engage a diverse representation of public experience, backgrounds, skills, and perspectives in the development, implementation and operations of the system.
- **Dignified experience:** does the person affected have a dignified experience, where they aren't required to overshare personal information, feel respected, are supported to succeed in their task, get a helpful service that anticipates their context and needs, etc?
- **User controls:** Do citizens/residents/etc have control over their own consents and data? Are data utilities available (like a means test Application Programmable Interface, or age verification service) to avoid unnecessary data sharing?
- **Data and decision provenance:** public sector organisations need to be able to trace, track and demonstrate provenance of data, decisions, rules and outcomes.
- **Whistleblowing:** Are there strong whistleblower protections, as an important last resort?
- **High veracity supply chains for government decisions systems** - establishing and managing a high veracity supply chain is no less important for service delivery or decision making than it is for physical goods. "Software/data supply chains" need high veracity assurance with appropriate obligations and monitoring of the performance of all actors in the value chain. This would help detect when things go wrong to detect, understand and respond.

# Recommendations

The following recommendations, combined with a traditional human review framework/mechanism, would help ensure an approach that is auditable, appealable, testable, human-centred and considered trustworthy, creating measurably ethical and fair government AI/ADM systems that benefit people.

Recommendations to **mandate/legislate**:

1. Establish a right of access by person to reasons of decision, in line with [Section 23 of the OIA in New Zealand](), creating a legal obligation on departments to provide explainability to the people about which decisions are made.
2. A requirement for immutably recording all decisions that impact a person, with explainability and references to legal authorities in real time, and a statement of reason (refer to the [Ombudsman's paper on automated decision making]()) for auditing and appeal.
3. Publish a publicly available reference implementation of high impact existing legislation/regulations as code, with a diverse range of test cases for reuse.
4. Establish requirements for where ML based systems **can't** be used in government, to avoid inconsistent outcomes and assure explainability where outputs are required to be contestable.
5. Establish real time monitoring of decisions, with patterns analysis & escalation mechanisms.
6. Establish public feedback mechanisms for AI/ADM systems, a public "bug reporting" tool.
7. Establish human-outcomes measurement framework (such as the [NSW Government Human Services Outcomes Framework]()) to incentivise systems and shape investment/prioritisation.
8. Monitor human outcome measures at budget, project, service and program levels to motivate net positive human outcomes by understanding and measuring for human impact (and harm), with escalation mechanisms. Harm in this sense, is when the human outcomes measures start to trend in the wrong direction, either at an individual, demographic or community level.
9. Establish a dedicated whole-of-government function, unconstrained by portfolio, for exploring, understanding and escalating issues that might put at risk public trust and confidence.
10. Include a requirement in drafting guidelines to draft all new regulations/legislation as human and machine readable from the start, to deliver clearer and better authoritative (human readable) legislation/regulation, and a reference implementation (machine readable) for reuse.
11. Establish public visibility of and participation in oversight for government AI/ADM systems.
12. Mandate the requirement to publish operational information and open source models, anonymised training data, algorithms and other relevant technical artefacts relating to gov ADM systems. This requires consideration about procurement/development requirements.
13. Establish a mandatory Government Service Standard that requires public services to measure and monitor human and policy impact alongside standard user and performance measures.
14. Map the end to end software, data and communication supply chains for critical digital services, systems and infrastructure, and establish high veracity mechanisms with monitoring
15. Ensure there are easy to access whistleblower and public reporting of issues or concerns.

Recommendations to **guide and support** government entities using AI/ADM:

16. Establish consistent and standardised rigorous testing of AI/ADM systems from different perspectives, ensuring the outcomes align with the expectations, and is representational.
17. Establish common requirements for quality management and end to end provenance for data.
18. Adopt the ALTAI framework for government use of AI/ADM and leverage the Algorithmic Impact Assessment as a means of determining the level of risk (which could leverage traditional models like the [ANAO risk framework]()), and design governance accordingly.
19. Establish public participation in policy development and service design as normal practice, independent from formal Communications and PR activities.
20. Establish an 'open by default' culture in public service, where 'need to know' is by exception.
21. Establish a culture of seeking and valuing feedback (with support in programme and resource planning) including end user feedback, and peer review from independent experts.

# Conclusion

The use of AI/ML and ADM systems in the public sector could provide a lot of value to government and society more broadly, but without clear guardrails and controls, could equally perpetuate and accelerate inequity and public distrust if used improperly. It is critical that public institutions carefully navigate this space and go well beyond the benchmark of minimum compliance, and towards best practice and exemplifying good public service through being lawful, ethical, accountable, fair and values-based. The use of AI/ML and ADM systems in the special context of government means these systems must stand up to the legal requirements and social expectations of a public sector to ensure the pursuit of genuine, accountable, just and equitable outcomes for all of the public.

# Further reading

Below are some related papers, frameworks, and additional considerations in this space:

- The Assessment List for Trustworthy AI (High Level Expert Group on AI, European Commission) https://altai.insight-centre.org/ with related ISO Standards https://www.iso.org/committee/6794475/x/catalogue/p/0/u/1/w/0/d/0 and an analysis of AI as to why and how the ISO work is fit to support the Act: https://op.europa.eu/en/publication-detail/-/publication/36c46b8e-e518-11eb-a1a5-01aa75ed71a1/language-en
- OECD recommendations on AI which form the basis for many multilateral discussions on AI https://legalinstruments.oecd.org/en/instruments/OECD-LEGAL-0449
- Algorithmic Impact Assessment framework from Canadian Government https://www.canada.ca/en/government/system/digital-government/digital-government-innovations/responsible-use-ai/algorithmic-impact-assessment.html
- Automated decision-making better practice guide by the Commonwealth Ombudsman, the Office of the Australian Information Commissioner and the Attorney-General's Department https://www.ombudsman.gov.au/publications/better-practice-guides/automated-decision-guide
- The OAIC 2020 Community Attitudes to Privacy Survey https://www.oaic.gov.au/engage-with-us/research/australian-community-attitudes-to-privacy-survey-2020-landing-page/2020-australian-community-attitudes-to-privacy-survey
- The Digital Council (NZ) paper on Trustworthy and Trusted ADM systems https://digitalcouncil.govt.nz/advice/reports/towards-trustworthy-and-trusted-automated-decision-making-in-aotearoa/
- The 2022 Edelman Trust Barometer
- The Australian Government policies and ethical frameworks for AI https://www.industry.gov.au/policies-and-initiatives/helping-industry-and-businesses-harness-technology/artificial-intelligence
- Building public trust in a landscape of artificial intelligence and digital technologies https://www.chiefscientist.gov.au/news-and-media/building-public-trust-landscape-artificial-intelligence-and-digital-technologies
- Citizens' engagement in policymaking and the design of public services (APH Library, 2011) https://www.aph.gov.au/about_parliament/parliamentary_departments/parliamentary_library/pubs/rp/rp1112/12rp01
- The Rise of Automated Decision Making in the Administrative State: Are Kerr's Institutions still 'Fit for Purpose'? https://www.auspublaw.org/blog/2021/08/the-rise-of-automated-decision-making-in-the-adminstrative-state-are-kerrs-institutions-still-fit-for-purpose

- *[Professional Computer Ethics and AI: House of Lords, Inquiry Governance and Virtue Ethics](https://works.bepress.com/mwigan/36/)*, Professor Marcus R Wigan, Monash University, (2018)
  https://works.bepress.com/mwigan/36/

# Appendix A: Applying the ALTAI to government

Below are the seven key requirements for trustworthy AI systems defined in the Assessment List for Trustworthy AI (**ALTAI**) framework, with considerations for each as they apply to the public sector use of AI/ADM/ML systems. The ALTAI is a useful framework, especially when combined with the special context of government, and should be considered the foundation of all public sector AI/ADM programs.

*"Human agency and oversight: AI systems should empower human beings, allowing them to make informed decisions and fostering their fundamental rights. At the same time, proper oversight mechanisms need to be ensured, which can be achieved through human-in-the-loop, human-on-the-loop, and human-in-command approaches"*

> The actions and decisions taken by public institutions should deliver measurable good public outcomes, whether enacted by a person or software. Both public servants and the public they serve should be empowered to make informed decisions towards net positive human outcomes, with human oversight.

*"Technical Robustness and safety: AI systems need to be resilient and secure. They need to be safe, ensuring a fall back plan in case something goes wrong, as well as being accurate, reliable and reproducible. That is the only way to ensure that also unintentional harm can be minimized and prevented."*

> The full AI system supply chain (software, data, rules, etc) should be secured and monitored, with non AI fallback systems available. AI system outputs should be explainable and testable against authoritative rules and/or use cases to ensure accuracy. Human impacts need to be measured (such as employment, public perceptions, homelessness, health, education, etc) so as to detect *some* forms of harm, to complement other qualitative measures of impact.

*"Privacy and data governance: besides ensuring full respect for privacy and data protection, adequate data governance mechanisms must also be ensured, taking into account the quality and integrity of the data, and ensuring legitimised access to data."*

> Governance can't simply be a self-management exercise. Departments should engage independent, expert participants onto AI governance boards to ensure rigour around review and oversight mechanisms. Data management, governance and oversight mechanisms need to be in place and continuously monitoring for quality, integrity and access management.

*"Transparency: the data, system and AI business models should be transparent. Traceability mechanisms can help achieving this. Moreover, AI systems and their decisions should be explained in a manner adapted to the stakeholder concerned. Humans need to be aware that they are interacting with an AI system, and must be informed of the system's capabilities and limitations."*

> Public institutions are especially accountable to the Parliament and general public, with scrutiny by the media, for all decisions and actions taken. Anything that affects a person or community should be explainable and immutably recorded, along with the data and traceability to the legal authorities of the action/decision. All government use of AI should communicate to the end user when they are interacting with an AI system, and given an option to pursue a human option if it is their preference.

*"Diversity, non-discrimination and fairness: Unfair bias must be avoided, as it could could have multiple negative implications, from the marginalization of vulnerable groups, to the exacerbation of prejudice and discrimination. Fostering diversity, AI systems should be accessible to all, regardless of any disability, and involve relevant stakeholders throughout their entire life circle."*

Government systems should be designed explicitly to avoid bias and ensure rule of law (where all are treated equally according to the law). This means even where training data is used, AI system outputs should be tested and testable against the law, and monitoring for bias must be proactive and actionable. All AI systems should be inclusively designed, with inclusive access to the resulting service.

*"Societal and environmental well-being: AI systems should benefit all human beings, including future generations. It must hence be ensured that they are sustainable and environmentally friendly. Moreover, they should take into account the environment, including other living beings, and their social and societal impact should be carefully considered."*

A human and environmental outcomes measurement framework is needed to ensure consistency of outcomes, to nudge AI systems towards achieving human and environmental outcomes (rather than purely financial outcomes, which can otherwise be at the cost of human outcomes) and to ensure government investment, services, budget and policy proposals, and project/program management are all reporting and delivering human and environmental outcomes.

*"Accountability: Mechanisms should be put in place to ensure responsibility and accountability for AI systems and their outcomes. Auditability, which enables the assessment of algorithms, data and design processes plays a key role therein, especially in critical applications. Moreover, adequate and accessible redress should be ensured."*

All government systems and services should have escalation mechanisms that are responsive to: change, unexpected trends/patterns, or negative human or environmental impacts (whether expected or not). AI Systems should be auditable (including legal traceability) and where a decision/action from an AI system affects a citizen/resident/community, the explanation should be provided in real time to the end user, along with accessible and easy to use appeals and feedback mechanisms. Feedback and appeals should feed back into the AI system design and operation.

# Appendix B: A trustworthy digital government checklist

Below is a useful checklist for helping assure trustworthy digital government programs, organised around the goals of accountability, human outcomes, AI/automation, and building ethically motivated teams. It is another perspective and layout for many of the ideas above.

## Traceability and accountability

- **Ensure, agree and document the principles and practices of Administrative Law** as it applies to digital systems to guide and drive the ethical and transparent use of digital, data and AI delivery across government.
- **Establish a [Better Rules](#) approach** for all new legislation and regulation, with publicly available reference implementations of all legislation and regulation as code.
- **Create an immutable record** of all decisions, based on what rules were invoked and with what authority to drive ease of auditing, records recall and visibility to citizens for their records.
- **Build in automated monitoring and escalation** of transactive services to ensure compliance with administrative law internally & externally.
- **Keep improving.** Develop active and continuous feedback loops from delivery back into policy/legislative improvement to provide for continuous improvement.

## Measurably good human outcomes

- **Don't ask for trust, build trustworthy systems together.** Engage with diverse communities to create measurement frameworks and to co-design policy, services and to ensure alignment of programs and delivery to public values and public good.
- **Build human-focused systems from the start.** Create and implement Government Service Standards that embed and normalise human outcomes and human measures of success.
- **Measure the outcomes that matter.** Proactively monitor quality-of-life outcomes at a process and line of business/ service scale, and link all activities to purpose, human outcomes and policy intent in a publicly accessible framework. Implement measurement frameworks such as the [NSW Human Service Outcomes Framework](#) across government, including in service delivery measures, budgets, business cases analysis, prioritisation frameworks, policy assurance.

## Be smart about AI and automation implementation

- **Get ready for machines to talk to machines.** Always assume machines will be users of your services, and plan for good and bad examples,
- **Measure the impact of algorithms.** Complement AI impact assessments with a measurement of human impact based on quality of life, and monitor for changes that indicate improved or worsening life indicators to identify harm. Actively plan for 'good' machine usage and mitigate 'bad' machines.
- **Assume humans cannot do it all by hand:** Technology scales. That also means when it breaks, it breaks at scale, and humans can't easily intervene. Include real-time monitoring for patterns in government services and policy interventions, to enable early identification of issues and escalation to the relevant people to address the problem, ideally before the system fails and certainly before the problem is widespread.

- **Prototype, test and scale.** Use agile, test driven & scalable techniques to create a policy-service spectrum that meets the evolving needs of the public, and that responds to continuous emergencies.

## Safe & ethically motivated teams and organisational structures

- **Thinking counts as work.** Make time in your programs and resourcing to think through hard problems, rather than settling on the most expedient approach. Simple tactics like building a ten percent innovation time factor into 'business as usual' operations are a good start.
- **Teach critical thinking skills.** Create a situational awareness of emerging trends and respond strategically in the interests of the public. Teach forecasting and critical thinking.
- **Choose good governance.** Evolve proactive and collaborative governance to empower service owners and ensure community and Country-centred design.
- **Develop 'by default' mindsets.** Pick your fundamental requirements—accessibility, trustworthiness, openness, and so on—and create systemic incentives that drive teams towards those outcomes by default.
- **Encourage candour.** Help teams build cultures that value peer review, transparency, public participation, and a sense of purpose.
- **Tie executive KPIs to human outcomes.** Include human measures in executive performance metrics, mandates, and reporting for agencies to help nudge good decisions.

# Appendix C: Concept diagram of trust mechanisms in government

Below is a concept diagram that proposes trust mechanisms for different parts of public institutions, to consider as another view on what is needed according to different roles of government.

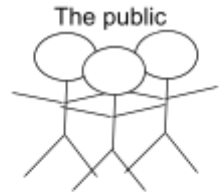

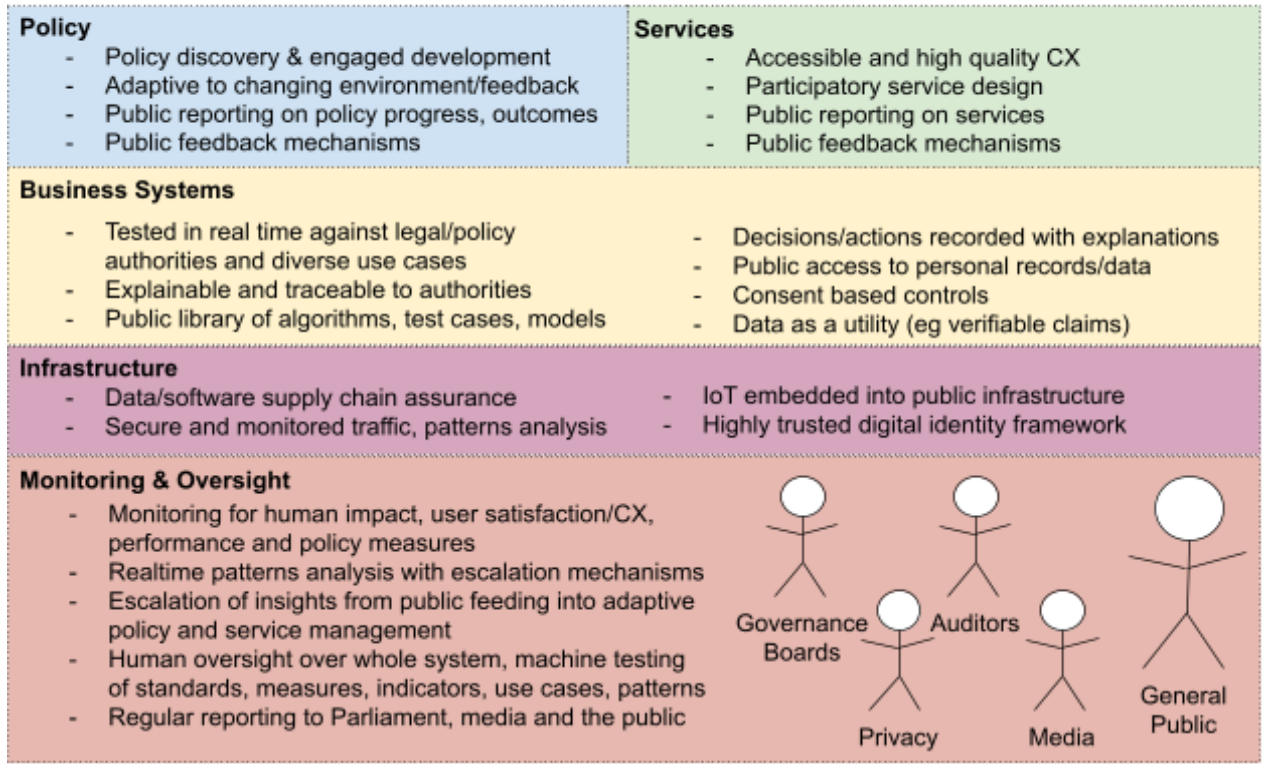

# Appendix D: Additional challenges to trust

Restoring public trust faces three big challenges: An industrial-era paradigm; the gaming of public opinion; and the opacity of automation.

## The industrial-era paradigm shift

*Despite plenty of digital transformation theatre, public institutions maintain centralised, need-to-know mindsets where citizens are treated as outsiders (or at best, "customers.") This leads to self-referential governance, and a pattern of seeking social licence to operate with impunity rather than engaging citizens in the process.*

The public sector generally seeks some form of permission from its denizens. However, it is common for a department to either assume or obtain public permission (usually in the form of existing or interpreted 'social licence'), then to act independently from the civil society it serves. The result is citizen disenfranchisement: People feel, and are, left out from the design, delivery and governance of the policies, decisions, and services that affect them. The public sector needs to do more than just deliver better services. It needs to create services and policies that are designed and delivered in a trustworthy way. That starts with more open, engaged, inclusive, equitable design processes where the public is acting alongside the government, instead of abdicating its needs to the public sector. Only by building *with*, rather than *for*, the public can we restore trust in public institutions.

## The gaming of public opinion

*The escalation of algorithmic misinformation and the rise of "parallel truths" have perverted many of the systems of policy-making, media coverage, and public engagement that once allowed governments to build consensus among the population.*

In the digital world, fakery can be indistinguishable from truth. When we can't verify a fact independently, we rely on our trust of the person or organisation delivering that information—we believe the messenger rather than the message. This is only going to get worse. Deepfake technology can automate the creation of believable videos of [anyone saying anything](), no matter how offensive, outrageous or worse, subtly misleading. This heralds a dark age in which saboteurs, criminals, trolls and bots act to game individuals, communities, and entire governments *en masse* for profit, crime — or just upvotes. The New Zealand Law Society commissioned an excellent [report into deepfakes in 2019](), which offered a range of regulatory recommendations worth considering domestically. But it also makes clear that the dominant threat will always be 'overseas', so such laws may not provide much protection. Robust institutions, citizen empowerment, and tools are needed.

## The opacity of automation

*Widespread adoption of technology as a replacement for paper processes, compounded by black-box AI decision-making, has given us impenetrable processes, systems that are hard to audit, and an inability to trace outcomes back to laws. This makes it hard for all citizens—particularly our most marginalised—to appeal, or even understand, the decisions of departments whose job is to serve them.*

A key factor in digital implementations is that they can be cheap and fast. Unlike physical products (such as a car) or in-person interactions (such as a service call), the marginal cost of another digital user is vanishingly small. That makes it very attractive for cash-strapped governments seeking to rein in spiralling budgets.

However, this is not without risk. As we digitise things, we automate them. As they become more complex, we apply machine learning and AI to them, relying on algorithms to triage and classify and respond and approve outcomes. In doing so, we make those processes opaque.

Government needs a better vision of what an 'augmented public sector' looks like. It must, above all else, have trust and real accountability designed in at the outset. Machines are fine for making services more responsive, proactive, and ubiquitous, but the aim should be to supplement and augment an empowered human workforce. Done right, AI can actually help *humanise* government services.

But as it stands, increased adoption of algorithms and automation often does just the opposite - mechanising services and marginalising citizens. Without strong and ethical service frameworks measured according to human outcomes—like the NZ Wellness Framework—we may miss an amazing opportunity to modernise public service in a way that gets humans and machines working together better than either can on its own.

# Appendix D: Considering the context of Digital Public Infrastructure

Below is a concept "Government as a Platform" framework (with layer-based principles) that provides a checklist of functions to explore, plan, secure and integrate for delivering services and public value. Consider what strategies your jurisdiction has in place for each capability listed.

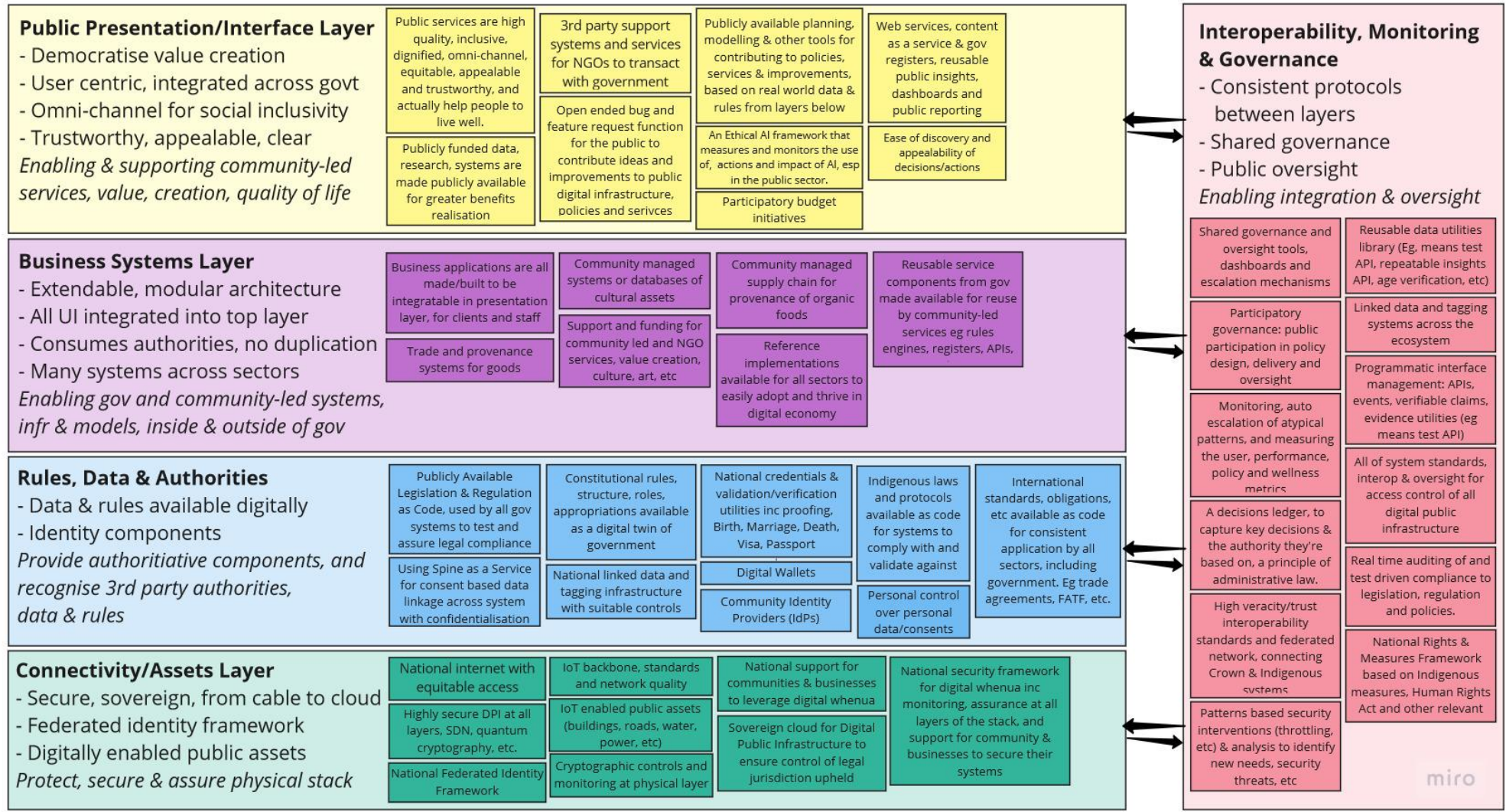